\documentclass[aps,prl,twocolumn,showpacs,preprintnumbers,amsmath]{revtex4}
\usepackage{amsmath}
\usepackage{epsfig}
\usepackage{array}
\usepackage{natbib}

\begin{document}

\title{A suggested search for $\boldsymbol{^{207}}$Pb nuclear Schiff moment in PbTiO$\boldsymbol{_3}$ ferroelectric}

\author{T.N. Mukhamedjanov, O.P. Sushkov}

\affiliation{School of Physics, University of New South Wales,\\
 Sydney 2052, Australia}

\begin{abstract}
We suggest two types of experiments, NMR and macroscopic magnetometry, with solid PbTiO$_3$ to search for the nuclear Schiff moment of $^{207}$Pb. Both kinds of experiments promise substantial improvement over the presently achieved sensitivities. Statistical considerations show that the improvement of the current sensitivity can be up to 10 orders of magnitude for the magnetometry experiment and up to 6 orders of magnitude for the NMR experiment. Such significant enhancement is due to the strong internal electric field of the ferroelectric, as well as due to the possibility to cool the nuclear-spin subsystem in the compound down to nanokelvin temperatures.
\end{abstract}


\maketitle

The existence of permanent electric dipole moment (EDM) of a quantum particle requires that fundamental parity ({\it P}) and time-reversal ({\it T}) symmetries are violated. By the {\it CPT}-theorem, this would also mean the violation of the combined {\it CP} (charge conjugation--parity) symmetry. Studies of {\it T} and {\it CP} violation in Nature provide valuable information for the theories of baryogenesis, and for our understanding of fundamental interactions in general. Thus, considerable effort has been put into searches for EDMs of particles, atoms and molecules.

The current experimental upper limit on the neutron EDM is $d_n \le 6.3 \times 10^{-26}\,e\ \textrm{cm}$ \cite{Neutron}.
Experiments with paramagnetic atoms and molecules, the most sensitive of which was performed with Tl atoms \cite{Com}, provide upper limit on the electron EDM, $d_e \le 1.6 \times 10^{-27}\,e\ \textrm{cm}$.
The most sensitive experiment with diamagnetic atoms is performed with $^{199}$Hg vapor \cite{Romalis}; it gives the upper limit on the EDM of $^{199}$Hg atom $d(^{199}\textrm{Hg}) \le 2.1 \times 10^{-28}\,e\ \textrm{cm}$.
This EDM is mainly induced by the nuclear Schiff moment, $S$, which is usually defined by the {\it P}- and {\it T}-odd electrostatic potential it generates \cite{SFK}:
\begin{equation}
\varphi(\boldsymbol{r})=4\pi(\boldsymbol{S}\cdot\boldsymbol{\nabla})\delta(\boldsymbol{r}) \ .
\end{equation}
Atomic calculations \cite{FKS,DFGK} show that the results of the $^{199}$Hg EDM experiment \cite{Romalis} lead to the following limit on the Schiff moment of the $^{199}$Hg nucleus.
\begin{equation}
S(^{199}\textrm{Hg}) < 0.75 \times 10^{-50}\,e\ \textrm{cm}^3=0.5\times10^{-25}\,e\ a_B^3
\label{HgSchiff}
\end{equation}
($a_B$ is the Bohr radius.)
Together with the neutron EDM data \cite{Neutron}, this result provides most significant limits
on the models of {\it CP}-violation in hadronic sector, see Refs.~\cite{KL,Romalis,DFGK}.

The EDM collaboration \cite{LANL} plans to improve the sensitivity of the neutron EDM measurements by two orders of  magnitude.
Comparable sensitivity is expected in the proposed deuteron EDM experiment \cite{BNL}.
The next generation experiments with the electron EDM are also on the way -- these are measurements with metastable levels of PbO  molecule \cite{PbO} and with YbF  molecule \cite{YbF}. The expected improvement of sensitivity compared to the atomic Tl experiment \cite{Com} is also about two orders of magnitude.
Romalis suggested to use liquid Xe for EDM measurements \cite{Xe}. Since Xe atoms are diamagnetic, the measurement is mainly aimed at the $^{129}$Xe nuclear Schiff moment.
By statistical considerations only, the sensitivity of the experiment is $d(^{129}\textrm{Xe}) \sim 3 \times 10^{-37}\,e\ \textrm{cm}$ for 10 days of averaging \cite{Xe}. Taking into account the smaller nuclear charge of Xe, this effectively means improvement by 8 orders of magnitude when compared to the present result (\ref{HgSchiff}) for $^{199}$Hg.
The limitations because of the systematic effects are also discussed in Ref.~\cite{Xe}.
The solid state experiment with gadolinium garnet, recently suggested by Lamoreaux \cite{Lam} to measure the electron EDM, promises 5 order of magnitude improvement over the current Tl result (statistical estimate corresponding to 10 days of averaging.)

In the present paper we suggest to use ferroelectric PbTiO${_3}$ to measure Schiff moment of $^{207}$Pb nucleus. Possibility to measure the macroscopic magnetization induced by electric field, as was suggested by Lamoreaux for gadolinium garnets \cite{Lam}, looks most promising because of the large internal electric field in ferroelectric. In addition, the nuclear-spin subsystem of the compound can be magnetically cooled down to nanokelvin temperatures. For a 10-day averaging, statistics allows to reach the sensitivity 10 orders of magnitude better than the present result (\ref{HgSchiff}).
Another possibility would be an NMR experiment ($^{207}$Pb is a spin-$\frac{1}{2}$ nucleus.) Although the sensitivity improvement is not as large in this case, the NMR experiment does not require nanokelvin temperatures.

PbTiO$_3$ is ionic crystal consisting of Pb$^{2+}$, Ti$^{4+}$, and O$^{2-}$ ions.
Although the internal electric field in ferroelectric crystal is very large, the average field acting on each charged particle in the compound is zero (the so-called Schiff theorem, \cite{Schiff}.)
It means, that calculating the EDM of the lead ion and multiplying it by the internal electric field is not, strictly speaking, the correct way to obtain the microscopic EDM Hamiltonian -- the lattice relaxes to compensate for the strong electric field, effectively screening it out.
The actual non-zero effect arises because of the finite size of ions and, in particular, due to the penetration of oxygen electrons inside the Pb ion in PbO$_{12}$ cluster (Pb$^{2+}$ ion and its first coordination sphere.)
In the ferroelectric state Pb ions are shifted with respect to the central position in the cluster by $X=0.47\ \textrm{\AA}$ \cite{Pb_disp}, this shift explains the strong ferroelectricity of PbTiO$_3$.
Wave functions of oxygen electrons penetrate inside the Pb ion, and, because of the displacement, create the electric field gradient at the Pb nucleus. Nuclear Schiff moment interacts with this gradient, which leads to the energy shift.

Similar effect has been considered previously for gadolinium garnets (Ref.~\cite{Simon}) where the ion displacement was induced by the external electric field. In the above paper, electronic properties relevant to the problem were described in the framework of GdO$_8$ cluster. In the case of PbTiO${_3}$ we have PbO$_{12}$ cluster, but for the first estimate this difference can, arguably, be neglected, and we apply the formula obtained in \cite{Simon} to our case:
\begin{eqnarray}
\Delta \epsilon/E_0 &\sim&b
\frac{Z^2}{(\nu_s \nu_p)^{3/2}}
\left(\dfrac{1}{3}R_{1/2}+\dfrac{2}{3}R_{3/2}\right)
\frac{(\boldsymbol{X}\cdot\boldsymbol{S})}{a_B \ e a_B^3}\ ,\nonumber\\
b&=&\frac{16}{\sqrt{3}}(\beta_s \beta_s^{\prime} + \beta_p \beta_p^{\prime})\ .
\label{de1}
\end{eqnarray}
Here $\Delta \epsilon$ is the energy shift caused by the Schiff moment of $^{207}$Pb nucleus, $\boldsymbol{S}$ \cite{footnote1};
$\boldsymbol{X}$ denotes the ion displacement, $E_0=27.2\ \textrm{eV}$ is the atomic energy unit, $e=|e|$ is the elementary charge,  $Z=82$ is the nuclear  charge of Pb; $R_{1/2}$ and $R_{3/2}$ are the usual relativistic enhancement factors. Dimensionless coefficients $\beta_s\sim-0.5$, $\beta'_s\sim 0.3$, $\beta_p\sim 0.7$, $\beta'_p\sim -0.2$
are related to the electronic structure of the crystal and allow to describe the $2p_{\sigma}$-electrons of the surrounding oxygen ions as the effective $s$- and $p$-electrons of the central lead ion with effective principal quantum numbers $\nu_s \sim 1.3$ and $\nu_p \sim 1.5$. A straightforward evaluation of Eq.~(\ref{de1}) gives the following energy shift:
\begin{equation}
\Delta \epsilon \sim -1.2\times 10^6 \ \frac{(\boldsymbol{X}\cdot\boldsymbol{S})}{a_B \ e a_B^3}\ \textrm{eV}
=-1.1\times 10^6 \ \frac{S}{e a_B^3}\ \textrm{eV} \ .
\label{de2}
\end{equation}
Note, that almost the same result can be obtained with the naive formula $\Delta\epsilon=d(\textrm{Pb}^{2+})\cdot E_{\textrm{int}}$, that is, completely ignoring the Schiff theorem for the ion. The internal electric field in the ferroelectric is about $E_{\textrm{int}} \sim 10^8\ \textrm{V/cm}$; having in mind
that the EDM of Pb ion is comparable to that of Hg atom, $d_{\textrm{Hg}}/(e\ \textrm{cm})\sim 10^{-2}S/(ea_B^3)$ \cite{FKS,DFGK},
we immediately arrive at the estimate (\ref{de2}). The two different approaches give very similar results because the ionic size is comparable to the separation between ions, and hence, the Schiff suppression for an ion is not as dramatic as it is for nucleus.

Taking the upper limit on the Schiff moment (\ref{HgSchiff}) as a reference point in Eq.~(\ref{de2}), we find the energy difference between the spin-up and spin-down nuclear states, essentially the shift of the $^{207}$Pb NMR line
\begin{equation}
2\Delta \epsilon \simeq 1.1\times 10^{-19}\ \textrm{eV};\ \ \Delta \nu =
\frac{2\Delta \epsilon}{h}\simeq 3\times 10^{-5}\ \textrm{Hz}.
\label{de3}
\end{equation}
The width of the NMR line is the central issue for any precise NMR measurement with the principal limitation coming from the dipolar broadening which cannot be removed by the spin echo technique. Since all the electron spins are compensated in the compound, the broadening is due to the magnetic dipole-dipole interaction of nuclear spins.
The second moment of the line shape $M_2=\int(\omega-\omega_0)^2P(\omega)d\omega$ is given by the following formula \cite{Slichter}:
\begin{equation}
M_2 = \frac{36}{5} \sum_j \left(\frac{\mu^2}{r_{j}^3}\right)^2
+\frac{16}{15}\sum_k \frac{I+1}{I}\left(\frac{\mu\mu'}{r_{k}^3}\right)^2,
\label{m2}
\end{equation}
$r_j$ and $r_k$ are the distances from a given $^{207}$Pb nucleus to all other nuclei with nonzero spin; the first summation is performed over the $^{207}$Pb sites (magnetic moment $\mu=0.59\mu_N$ \cite{tabisot}), the second summation is over the following sites:
$^{47}$Ti (spin $I=5/2$, magnetic moment $\mu'=-0.79\mu_N$),
$^{49}$Ti ($I=7/2$, $\mu'=-1.10\mu_N$), and
$^{17}$O ($I=5/2$, $\mu'=-1.89\mu_N$).
For the natural abundance of isotopes,  22.1\% $^{207}$Pb, 7.4\% $^{47}$Ti, 5.4\% $^{49}$Ti, and 0.038\% $^{17}$O, Eq.~(\ref{m2}) gives $\sqrt{M_2}=4.1\times 10^{-13}\,\textrm{eV}$.
Assuming the Gaussian shape for the NMR line, the half-width is
\begin{equation}
\Gamma =\sqrt{8\ln{2}\,M_2}\simeq 9.6\times 10^{-13}\ \textrm{eV}, \ \ \
\Delta\nu_{\Gamma}=\frac{\Gamma}{h}=230\ \textrm{Hz} \ .
\label{g}
\end{equation}
Interaction with magnetic lead nuclei accounts for 57\% of the total value of $M_2$, interaction with $^{47,49}$Ti gives 40\%, and $^{17}$O gives about 3\% of $M_2$.
The Pb NMR data for ceramic PbTiO$_3$ is available \cite{Korn}.
Authors of the above paper claim that their data is in agreement with the estimate according to formula (\ref{m2}).

It is interesting to compare our estimates (\ref{de3},\ref{g}) with the parameters of the $^{199}$Hg experiment \cite{Romalis}.
The effect (\ref{de3}) is a factor $\sim 0.5\times 10^5$ larger than that for atomic Hg, but the line-width (\ref{g}) is also larger by factor $\sim 2\times 10^5$.
The number density of $^{207}$Pb in the compound at the natural abundance is
$n\approx 0.33\times 10^{22}\ \textrm{cm}^{-3}$, while the number density of $^{199}$Hg in the
experiment \cite{Romalis} was about $n \sim 10^{14}\ \textrm{cm}^{-3}$.
Assuming equal volumes, full polarization of $^{207}$Pb nuclei, and assuming that sensitivity scales as $ \propto  \Delta\nu\sqrt{n/\Gamma}$,
we find that sensitivity to the Schiff moment can be improved by 6 orders of magnitude, compared to the atomic Hg experiment \cite{Romalis}.

It is possible to reduce the dipole-dipole broadening by using
samples grown to have reduced concentration of magnetic
isotopes. Removal of $^{47}$Ti and $^{49}$Ti should reduce the
width (\ref{g}) by 30\%. Further reduction is possible in the
sample with depleted concentration of $^{207}$Pb.
The second moment $M_2$ scales linearly with  the number density of $^{207}$Pb. 
In the sufficiently depleted sample the shape of the line
is not Gaussian and while $\sqrt{M_2}\propto \sqrt{n}$, the
line-width should actually scale as $\Gamma \propto n$.
The reason for this difference in scaling is as follows: the value of $M_2$ is determined by the rare events of two $^{207}$Pb nuclei occupying neighboring sites; the value of $\Gamma$, however, is mainly determined by the typical events of $^{207}$Pb nuclei separated by the average distance $r\sim n^{-1/3}$.
Therefore, for example, at $n \sim 10^{19}\,\textrm{cm}^{-3}$ the line-width is
$\Delta\nu_{\Gamma}\sim 1\ \textrm{Hz}$.
The width reduction at the expense of the number density does not improve the statistical sensitivity,
but it can be helpful for analysis of systematics.

The spin-lattice relaxation is probably not a serious issue. For the
sample with the natural abundance of isotopes the corresponding
lifetime is about 1~sec \cite{Kim}. We attribute this relaxation
to the combined action of magnetic interaction between $^{207}$Pb
and $^{47,49}$Ti nuclei and electric quadrupole interaction of
$^{47,49}$Ti nuclei with the lattice. If this is correct, the
spin-lattice relaxation time in the sample with no $^{47,49}$Ti
nuclei must be much larger than 1~sec.


Because of the substantial hysteresis in the ferroelectric, the energy dissipation in experiments with the external electric field reversal can be a serious issue. On the other hand, in the experiment suggested here there are no external electric field reversals. The external electric field creates a residual polarization in the ferroelectric and is then switched off. The polarized sample is then rotated, effectively reversing the internal electric field.

Another approach, which looks even more promising than the NMR one, is to measure the magnetization induced by the external electric field, as was suggested by Lamoreaux \cite{Lam} to measure the electron EDM in gadolinium garnets. Compared to the electron EDM case, we immediately lose 3 orders of magnitude just on the nuclear magneton vs the Bohr magneton ratio. However, the effective electric field in ferroelectric is 4 orders of magnitude larger than the external electric field in the gadolinium garnet experiment. Additionally, it should be possible to cool the nuclear spins in PbTiO${_3}$ down to $10\ \textrm{nK}$, whereas the lowest experimental temperature in gadolinium garnet is probably limited to about $1\ \textrm{K}$ (in both cases, the lowest temperature is determined by the spin freezing); the last argument wins about 8 orders of magnitude in sensitivity.

Let us perform more accurate estimates. $^{207}$Pb nuclear magnetization induced by the Schiff moment is
\begin{equation}
M = n\mu\frac{\Delta \epsilon}{k_BT}=-1.1\times 10^6 \ n\mu \ \frac{S}{e a_B^3}\frac{1\ \textrm{eV}}{k_BT} \ .
\label{magn}
\end{equation}
Here $n$ is the number density of $^{207}$Pb nuclei, $\mu$ is their magnetic moment; $\Delta \epsilon$ is
given by Eq.~(\ref{de2}).
In metals, temperatures below $1\ \textrm{nK}$ have been reported to be achieved by magnetic cooling of spins (e.g., see review \cite{Cool}.)
Magnetic cooling in insulators is not as developed; however, the temperature $T\sim 10^{-6}\,\textrm{K}$ has been achieved in the studies of nuclear magnetic ordering in CaF$_2$ \cite{Cool}. Nuclear spins in PbTiO$_3$ freeze at sufficiently lower temperature, forming a spin glass state.
The freezing  temperature $T_f$ is determined by the spin-spin interaction, therefore,
\begin{equation}
\label{fr}
T_f \sim \frac{\Gamma}{k_B} \sim 10^{-8}\,\textrm{K} \ ,
\end{equation}
where $\Gamma$ is given in (\ref{g}).
We take this temperature as the lowest possible for the experiment and substitute $T=T_f=10\ \textrm{nK}$ in Eq.~(\ref{magn}). The expected magnetic induction is then
\begin{equation}
B = 4\pi M= \frac{S_N}{e a_B^3}
\cdot 1.8\times 10^{17}\ \textrm{G}\ .
\end{equation}
According to Ref.~\cite{Lam}, one can achieve the sensitivity of $3\times 10^{-16}\ \textrm{G}$ for 10 days of averaging with a SQUID magnetometer, and it is even possible to do 2 orders of magnitude better with magnetometry based on nonlinear Faraday effect \cite{Budker,Lam}.
The corresponding sensitivity to the nuclear Schiff moment is
\begin{equation}
S \sim 1.7\times 10^{-35}ea_B^3=2.5\times 10^{-60}\,e\ \textrm{cm}^3,
\label{10}
\end{equation}
which is 10 orders of magnitude better than the present result
(\ref{HgSchiff}) for $^{199}$Hg. The sensitivity (\ref{10}) is 4
orders of magnitude better than the value of the nuclear Schiff
moment predicted by the standard model, $S\sim 10^{-56}\,e\
\textrm{cm}^3$, see Refs.~\cite{SFK,FKS,He,KL}.


Similarly to the NMR experiment discussed above, energy dissipation from the external electric field reversals can be a serious issue. This problem can be avoided by switching off the external electric field altogether, and rotating the electrically polarized sample to effectively reverse the internal electric field. With this kind of ``rotational reversal'' paramagnetic impurities (electronic paramagnetism) can be a source of potentially dangerous systematics, so the sample should be prepared to be free of such impurities. Nonmagnetic impurities do not contribute to these systematics.

To summarize, we have suggested two types of experiments, NMR and macroscopic magnetometry, in solid PbTiO$_3$ to search for the nuclear Schiff moment of $^{207}$Pb. Both kinds of experiments promise substantial improvement over the presently achieved sensitivities. Statistical considerations show that the magnetometry experiment can offer up to 10 orders of magnitude improvement over the present sensitivity to the nuclear Schiff moment ($^{199}$Hg experiment \cite{Romalis}.) With this improvement, the sensitivity is 4 orders of magnitude better than the value of the nuclear Schiff moment expected from the standard model, \cite{SFK,FKS,He,KL}.
Such significant enhancement is due to the strong internal electric field of the ferroelectric, as well as due to the possibility to cool the nuclear-spin subsystem in the compound down to nanokelvin temperatures.
The NMR-type experiment does not require such low temperatures, but offers smaller improvement in sensitivity -- about 6 orders of magnitude when compared to the $^{199}$Hg experiment \cite{Romalis}.

We are grateful to  M.V. Romalis, J.M. Cadogan, J. Haase, A.O. Sushkov, R. Newbury, I.B. Khriplovich,
D. Budker, and S. K. Lamoreaux for very important discussions and comments.

\end{document}